\documentclass[prb,twocolumn,floatfix,showpacs]{revtex4}
\usepackage{graphicx,subfigure,amsmath,colordvi,times}
 \usepackage[usenames]{color}
 \usepackage{multirow}
\usepackage{ulem}
\usepackage{epstopdf}

\newcommand{\be}{\begin{equation}}
\newcommand{\ee}{\end{equation}}
\newcommand{\bes}{\begin{equation*}}
\newcommand{\ees}{\end{equation*}}
\newcommand{\bea}{\begin{eqnarray}}
\newcommand{\eea}{\end{eqnarray}}
\newcommand{\bean}{\begin{eqnarray*}}
\newcommand{\eean}{\end{eqnarray*}}
\newcommand{\ba}{\begin{array}}
\newcommand{\ea}{\end{array}}
\newcommand{\kv}{{\bf k}}

\begin{document}
 \title{Using magnetic stripes to stabilize superfluidity in electron-hole double monolayer graphene }
\author{Luca Dell'Anna$^1$, Andrea Perali$^2$, Lucian Covaci$^3$, and David Neilson$^{2}$}
 \affiliation{
$^1$ Dipartimento di Fisica e Astronomia ''G. Galilei'' and CNISM, Universit\`a di Padova, 35131 Padova, Italy\\ 
$^2$ Dipartamenti di Fisica e di Farmacia, Universit\`a di Camerino, 62032 Camerino, Italy\\
$^3$ Department of Physics, Universiteit Antwerpen, Groenenborgerlaan 171, 2020 Antwerpen, Belgium} 

\begin{abstract}
Experiments have confirmed that double monolayer graphene does not generate finite temperature electron-hole superfluidity, because of very strong screening of the pairing attraction. The linear dispersing energy bands in monolayer graphene block any attempt 
 to reduce the strength of the screening. We propose a hybrid device with two sheets of monolayer graphene in a
 modulated periodic perpendicular magnetic field.  The field preserves the isotropic Dirac cones of the original monolayers but  reduces the slope of the cones, making the monolayer Fermi velocity $v_F$ smaller.  We demonstrate that with current experimental techniques, the reduction in  $v_F$ can  weaken the screening sufficiently to allow electron-hole superfluidity at measurable temperatures.  

\end{abstract}
\maketitle

\section{Introduction}

The  transition temperatures for electron-hole superfluidity in thin parallel conducting sheets of  electrons and holes are expected to be high because 
the electron-hole pairing is Coulombic and is strong compared with conventional superconductors.  This has led to suggestions of 
superfluidity at room temperatures in double electron-hole monolayers of  graphene  \cite{MacDonald2008}, but 
Ref.\ \onlinecite{Lozovik2012} showed that strong screening of the pairing attraction in this system tends to suppress finite temperature 
superfluidity.   Here we propose a hybrid double monolayer graphene device designed to boost the 
pairing attraction by reducing the effects of the screening, and we demonstrate that this can lead to magnetically 
induced superfluidity. We use electronic band structure engineering, coupling periodic real or pseudo magnetic fields 
to double electron-hole monolayers of  graphene separated by a thin insulating barrier.  
The resulting quantum properties of the device stabilize macroscopic quantum coherence 
and allow, in a solid state electronic device, the tuning of the strength of the many-body correlations and the related superfluid 
properties. Previously, such tuning has only been possible in ultra-cold 
fermionic atoms \cite{coldatoms}.  

Both theory \cite{Lozovik2012,Perali2013,Germash} and experiment \cite{Gorbachev2012} have 
established that conventional electron-hole double monolayer graphene does not generate  finite temperature electron-hole 
superfluidity because {of strong} screening of the electron-hole pairing.  
This originates from the linear 
Dirac cones of  the monolayer graphene  band structure, $\epsilon_\pm(\mathbf k)=\pm\hbar v_F |{\mathbf k}|$, with constant
Fermi velocity $v_F$, that makes the 
Fermi energy $E_F$ dominate the average Coulomb interaction
$\langle V_{Coul}\rangle$.
The resulting small  interaction strength parameter, $r_s=\langle V_{Coul}\rangle/E_F= e^2 /(\hbar v_F \kappa)$ that is fixed independent of density, leads to strong screening that makes    
pairing too weak for finite-temperature superfluidity to occur \cite{Lozovik2012}.  Only when  $r_s > r_s^\textrm{onset} = 2.3$ does the 
pairing become sufficiently strong for a large superfluid gap $\Delta$ to open discontinuously and suppress  the screening.  
 With graphene on a hexagonal boron-nitride (h-BN) substrate of dielectric constant $\kappa \simeq 3$$-$$4$ ,  $r_s< 1$.  
Theoretically, weak-coupled superfluidity could still occur, but at impractically low temperatures.  Since it would  be destroyed by residual disorder \cite{Abergel} we do not consider it further.
Other systems  have  been proposed for observing the superfluidity with an $r_s$ parameter that can be varied with the density.  These include two  sheets of  multilayer graphene which have non-linear  dispersing energy bands \cite{Perali2013,Neilson2014,Zarenia2014}, double quantum wells in 
GaAs \cite{Croxall,Lilly}, and hybrid GaAs--graphene structures \cite{Pelligrini}.
	
In this paper we propose use of a periodic magnetic field  applied perpendicular to double electron-hole  monolayer graphene in order to reduce the slope of the monolayer
Dirac cones while preserving their isotropy \cite{6,7,8,9}.  This reduces $v_F$ 
and increases the value of $r_s$.  If $r_s$ can be increased to $r_s>r_s^\textrm{onset}$, then 
finite-temperature superfluidity can occur \cite{Lozovik2012}.
The Fermi velocity in monolayer graphene can also be renormalized, 
but  non-isotropically, by applying a one-dimensional 
potential superlattice in the layer \cite{Barbier}.  

\section{Methods}
\subsection{Reducing the Fermi velocity} 

\begin{figure}[t]
\includegraphics[height=4.5cm,width=0.4\textwidth]{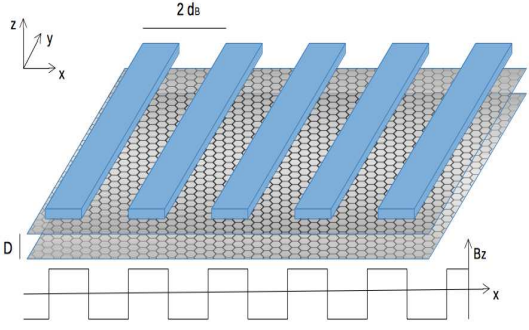}
\caption{(Color online) Possible realization of the device. An array of ferromagnetic stripes with periodicity $2d_B$ placed on top of two monolayer sheets of graphene, produces a periodic magnetic field  $B_z \simeq \pm B$.}  
 \label{Fig.1}
 \end{figure}

Consider a magnetic field in the $z$-direction, perpendicular to the monolayers, as a periodic array in the 
$x$-direction of rectangular magnetic barriers and wells of height $B_z = \pm B$ and width $d_B$.
The field could be generated with a periodic array of ferromagnetic stripes placed 
on top of the graphene monolayers (Fig.\ \ref{Fig.1}).

Since the average of the magnetic flux  is zero across the unit cell of the periodic field,
the main effect of the  field is  to modify the  monolayer band structure. 
With zero flux, the results are insensitive to fine details of the magnetic profile \cite{8}.  
At low energies the de Broglie wavelengths of the
quasiparticles are much longer than the length scale of the magnetic
field variation, so the magnetic profile can be approximated by a periodic square wave 
magnetic field \cite{3}.
We assume  smearing of the magnetic barriers is much 
greater than the lattice spacing. The smoothness of the vector 
potential on a microscopic scale means we can neglect intervalley 
scattering and use  single-valley continuum Dirac-Weyl theory. 
The Zeeman effect is very small in graphene, and electron-hole pairing is insensitive to relative spin orientation,  so
we can neglect spin effects induced by the magnetic field.

With this magnetic profile, the vector
potential in the Dirac Hamiltonian can be fixed by the Landau gauge and chosen periodic in the $x$ direction. 
The spectrum can then be obtained by a standard transfer matrix approach 
\cite{mckeller}: Define ${\cal T}$ (depending on $B$, momentum $k_y$ and 
energy level $\epsilon$) as the transfer matrix that relates 
the two-component wavefunction on $x$ to its value on $x\pm 2d_B$.
Showing that $\det {\cal T}=1$ and introducing a new momentum $k_x$ due to the 
periodicity of the superlattice,  the eigenvalues of ${\cal T}$ can  be written 
as $e^{\pm 2ik_x d_B}$, so the condition determining the band structure is
\be
\label{TrT}
\textrm{Tr}\left[{\cal T}(k_y,\epsilon)\right]=2\cos(2k_x d_B).
\ee
Expanding $\textrm{Tr}\left[{\cal T}(k_y,\epsilon)\right]$ on $k_y$ and $\epsilon$,
noting that $\textrm{Tr}\left[{\cal T}(k_y,0)\right]=2\cosh(2k_yd_B)$, and expanding $\cos(2k_x d_B)$ on $k_x$, we then solve Eq.\ (\ref{TrT}) for $\epsilon$.  We find that the energy dispersion 
remains linear and isotropic for small momentum, but with a reduced velocity $\alpha_d v_F\le v_F$ 
\cite{6,7,8,9}, 
\be
\epsilon_\pm(\kv)=\pm \hbar( \alpha_d v_F) |\kv| \left(1+ \delta(\kv)\right)\ . 
\label{epsilonk}
\ee
In Eq. (\ref{epsilonk}) $\alpha_d$ is a function of $d=d_B/\ell_B$, the dimensionless stripe width, where the magnetic length 
$\ell_B=\sqrt{\hbar c/eB} \simeq 26/\sqrt{B (\mathrm{T})}$ nm.      
An expansion of Eq. (\ref{TrT}) limits the correction term in Eq. (\ref{epsilonk}), 
$|\delta(\kv)|\lesssim d_B^2k_x^2/6$.
The decrease in Fermi velocity in the  small and large $d$ limits is \cite{6,9},   
\bea
&&\alpha_d\simeq 1-d^4/60,  \ \ d\ll 1 \\ %\ ;\ \ \ \ \ \  
&&\alpha_d\simeq \frac{2d}{\sqrt{\pi}}e^{-d^2/4},  \ \  d\gg 1 \ .
\label{alpha_dlarge}
\eea

At large densities, $E_F$  eventually passes out of the first energy band of the periodic magnetic field into the band gap where the linear spectrum approximation is no longer valid. 
The energy width $E_B$ of the band can be numerically 
calculated by solving  Eq.\ (\ref{TrT}) at the boundaries   
of the Brillouin zone, $k_x=\pm\pi/2d_B$, i.e. 
$\textrm{Tr}[{\cal T}(0,E_B)]=-2$, with $E_0=\hbar v_F/\ell_B \approx 22\sqrt{B\text{ (Tesla)} }$ meV.   
We find $E_0 \alpha_d/d\le E_B < E_0 \pi\alpha_d/2d$.
The lower limit is valid for large $d$.  It corresponds to the spectrum along $k_x$ as $\epsilon(k_x)=\hbar(\alpha_d v_F)\sqrt{[1-\cos(2k_x d_B)]/2}/d_B\simeq \hbar(\alpha_d v_F)k_x(1-d_B^2k_x^2/6)$, the largest 
deviation from linearity.  The upper limit is  valid for small $d$.  It corresponds to a completely linear spectrum, $\epsilon(k_x)=\hbar\alpha_d v_F k_x$. 
Remarkably, along the $k_y$ direction the spectrum is found to be almost linear for all $\epsilon \le E_B$ and for all parameters we  use.
Within linear approximation, we must restrict our results to  values of
$E_F\le E_B$.   
Even in the least favourable case, when 
the extreme value $E_F=E_B$ is reached at $k_{x}=1/d_B$ (in the linear approximation), the true spectrum remains close to the linear spectrum, $\epsilon(\frac{1}{d_B})=\sqrt{(1-\cos{2})/2}E_F \simeq 0.84 E_F$.  This gives an estimate of the maximum error along $k_x$.
In the superfluid state, $\delta(\kv)$
must also be small compared with the energy gap $\Delta_{\mathrm{max}}$
since the gap excludes single-particle states lying less than  $\Delta_{\mathrm{max}}$ above $E_F$, or equivalently,  
using the limiting value at the  
Brillouin zone edge, $|\delta(\kv)|<\pi^2/24$,  
\be
\pi^3\alpha_d/48 d \ll \Delta_{\mathrm{max}}/E_0\ .
\label{devonDelta}
\ee
For $d\gg 1$, Eq.\ (\ref{devonDelta}) is always satisfied since $\alpha_d\sim d\mathrm{e}^{-d^2/4}$.

\subsection{Enhancement of superfluidity by magnetic field}

The effective Hamiltonian for the two monolayer sheets of graphene in the presence of the periodic perpendicular magnetic field is,
\begin{eqnarray}
{\cal{H}}\! =\! \sum_{\mathbf{k}\gamma} \xi_{\mathbf{k}}^{\gamma} c^{\gamma \dagger}_{\mathbf{k}} c_{\mathbf{k}}^{\gamma}
 \!+\!\sum_{\substack{\mathbf{q}\mathbf{k}\gamma\\ \mathbf{k}'\gamma'}} V^{eh}_{\mathbf{k}-\mathbf{k}'}
c^{\gamma \dagger}_{\mathbf{k}+\frac{\mathbf{q}}{2}}
c^{\gamma  \dagger}_{-\mathbf{k}+\frac{\mathbf{q}}{2}}
c^{\gamma' }_{\mathbf{k}'+\frac{\mathbf{q}}{2}}
c^{\gamma' }_{-\mathbf{k}'+\frac{\mathbf{q}}{2}}\ .
\label{Grand-canonical-Hamiltonian}
\end{eqnarray}
The single-particle energy bands for the modified Dirac cones of the conduction band (electrons) 
and valence band (holes),  
$\xi_{\mathbf{k}}^{\gamma}= \gamma \alpha_d v_F |{\mathbf{k}}| - \mu$, 
are measured from their respective chemical potentials $\pm \mu$,
where  $\gamma=1 (-1)$ labels the electron (hole) sheet.  
The $c^{\gamma \dagger}_{\mathbf{k}}$ and $c_{\mathbf{k}}^{\gamma}$
are creation and destruction operators for  electrons and holes.  Spin indices are implicit. 
$V^{eh}_{\mathbf{q}}$ is the screened electron-hole interaction.

The mean-field equations at zero temperature for the momentum-dependent superfluid gap functions 
$\Delta_{\mathbf{k}}^{\gamma}$, and for equal electron and hole densities $n_{+}=n_{-}=n$ are,
\begin{eqnarray}
\Delta_{\mathbf{k}}^{\gamma}
 &=& - \frac{1}{\Omega}\sum_{\mathbf{k}'\gamma'}
F^{\gamma\gamma'}_{kk'}%\frac{1}{2}
V^{eh}_{\mathbf{k}-\mathbf{k}'}
\frac{\Delta_{\mathbf{k}'}^{\gamma'}
}{2 E_{\mathbf{k}'}^{\gamma'}}
\label{Delta-eqn} \\
n_{\gamma}
&=& \frac{g_v g_s}{\Omega} \sum_{\mathbf{k}}\frac{1}{2}\!
\left(1-\frac{\xi_{\mathbf{k}}^{\gamma}}{E_{\mathbf{k}}^{\gamma}} \right) \ .
\label{n-eh}
\end{eqnarray}
$E_{\mathbf{k}}^{\gamma} =\sqrt{\xi^{\gamma\ 2}_{\mathbf{k}} +\Delta_{\mathbf{k}}^{\gamma\ 2}}$,  $g_s(g_v)=2$ are the spin 
(pseudospin) factors, and 
$\Omega$ is the sheet area.  
We retain only the s-wave harmonic in the graphene form factor $F^{\gamma\gamma'}_{kk'}=1/2$ which 
comes from the overlap of the 
single-particle wave-functions in the strong-coupled regime \cite{Lozovik2012}.

We self-consistently calculate the screened electron-hole interaction 
$V^{eh}_{\mathbf{q}}$ within the random phase approximation (RPA) in the 
zero temperature superfluid state \cite{Lozovik2012,Perali2013,MacDonald2012}.
The most favourable conditions for pairing are at small interlayer 
separations $D$ on the scales of both the effective Bohr radius and the inverse Fermi momentum $k_F^{-1}$ in each layer. 
In this case $qD\ll 1$ and {the interaction} reduces to,
\begin{equation}
\!V^{eh}_{\mathbf{q}}\!=\!\frac{v_q\mathrm{e}^{-qD}}{1+2v_q\Pi(q)}\ .
\label{Vscr}
\end{equation}
$v_q=-2\pi e^2/(\kappa q)$ is the unscreened Coulomb interaction.  
$\Pi(q) = \Pi^{(n)}(q)+ \mathrm{e}^{-qD}\Pi^{(a)}(q) \simeq \Pi^{(n)}(q)+\Pi^{(a)}(q)$ is the sum of the normal (intralayer) and anomalous (interlayer) 
polarizabilities for the superfluid state calculated with the linear energy spectrum.
The gap equation [Eq.\ (\ref{Delta-eqn})] is independent of density when expressed in units of $E_F$ and $k_F$, 
with the exception of the $\mathrm{e}^{-qD}$ factor in Eq.\ \ref{Vscr} for $V^{eh}_{\mathbf{q}}$.  With increasing density, this factor weakens $V^{eh}_{\mathbf{q}}$.
\begin{figure}[t]
\includegraphics[height=5.5cm,width=0.48\textwidth]{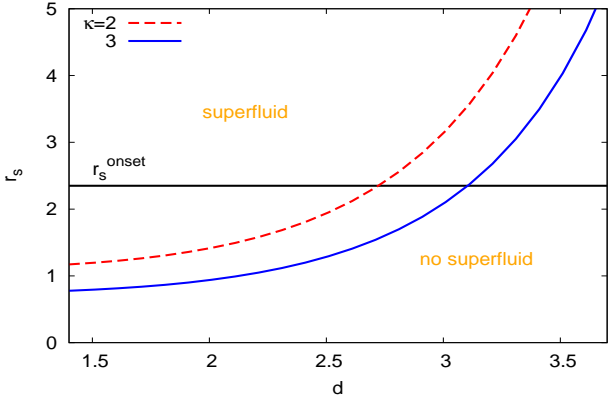}
\caption{(Color online) Interaction parameter $r_s$ for magnetic field of periodicity $2d$ for dielectric constant  $\kappa=3$ (monolayers embedded in a h-BN substrate and  
$\kappa=2$ (free standing monolayers, separated by h-BN).
Superfluidity occurs for 
$r_x\ge r_s^\textrm{onset}= 2.35$ \cite{Lozovik2012}.}
\label{rs-alpha}
\end{figure}
Rescaling $\tilde{q}=\alpha_d q$ gives $\Pi(q,\alpha_d v_F)=\alpha_d^{-2}\Pi(\tilde{q},v_F)$, and, for  $\tilde{q}D \ll 1$, 
$V^{eh}_{\mathbf{q}}(\alpha_d v_F,\kappa)= \alpha_d^2 V^{eh}_{\mathbf{q}}(v_F,\alpha_d\kappa)$.
Equations \ref{Delta-eqn} and \ref{n-eh} thus remain the same, but with $(\alpha_d \kappa)$ replacing $\kappa$.
The interaction parameter $r_s= [1/\alpha_d][{e^2}/({\hbar \kappa v_F})]$ increases by a factor $\alpha_d^{-1}$.
  
\section{Results}
  
Figure \ref{rs-alpha} shows that by tuning $\alpha_d$ with $d$ using a periodic 
magnetic field, $r_s$ can be increased above  the value $r_s^\textrm{onset}$ needed for superfluidity. 
$\kappa=3$ corresponds to  monolayers embedded in a h-BN substrate.   
$\kappa=2$ corresponds to a free standing system with the two monolayers separated by h-BN.
For $d<d_{\textrm{min}}\approx 2\sqrt{\log \kappa}+1$, $r_s<r_s^\textrm{onset}$, and  
the superfluidity is killed by strong screening. For $\kappa=3(2)$, 
$d_{\textrm{min}}\approx 3.1(2.7)$.

The renormalization of the band structure is the main effect that 
drives the  superfluidity in the double monolayer 
 system.  We find whenever superfluidity occurs, the electron-hole pairs of the 
superfluid ground state are compact compared with their spacing, 
making them approximately neutral, and the electrons and holes have opposite wave-vectors.  
Thus screening effects and  effects of the  
magnetic field on the orbital degrees of freedom should be small compared 
with the primary effect,  the renormalization of the Fermi velocity. 

\begin{figure}[t]
\includegraphics[height=5.5cm,width=0.48\textwidth]{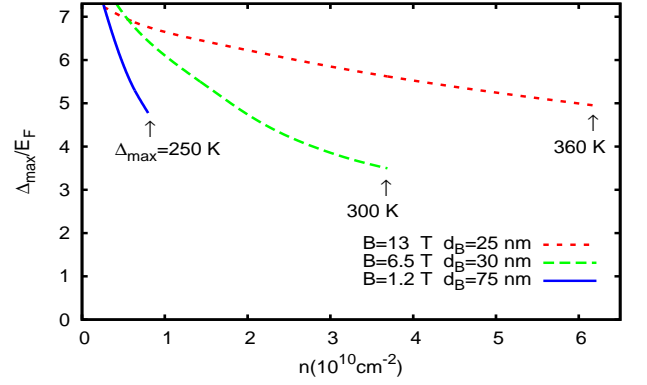}
\caption{(Color online) Maximum superfluid gap $\Delta_{\mathrm{max}}$  for different values of the magnetic field $B$, as a function of 
sheet densities $n$. Sheet separation  is $D=2$ nm. At arrow points: $\Delta_{\mathrm{max}}$  in Kelvin.}
\label{Delta}
\end{figure}

Figure \ref{Delta} shows the maximum superfluid energy gap $\Delta_{\mathrm{max}}$ at zero temperature  for different values of the magnetic field $B$,  
as a function of sheet density $n$, for  two monolayers separated by $D=2$ nm and embedded in a h-BN substrate.
The gaps $\Delta_{\mathrm{max}}$ are 
of the order of several hundred Kelvin.
$\Delta_{\mathrm{max}}$ decreases with increasing $n$, due to the  $\mathrm{e}^{-qD}$ factor in $V^{eh}_{\mathbf{q}}$ [Eq.\ \ref{Vscr}].   
This eventually results in no solution to the gap equation.  However, 
we terminate the curves in Fig.\ \ref{Delta} when $E_F$ reaches $E_B$, and this occurs before such a density is reached. 
$E_F=E_B$ thus gives a lower limit on the maximum density for the superfluid phase,  $n\simeq 5\times 10^{10} B \mathrm{(T)}/d^2$ cm$^{-2}$.

For all the gaps shown in Fig.\ \ref{Delta}, $\Delta_{\mathrm{max}} \gg E_F$, leading to a strong suppression of the screening.   At higher densities, the very strong screening  would kill the superfluidity before the system can enter the BCS regime \cite{Perali2013}.  However, with increasing 
density, and well  before $\Delta_{\mathrm{max}}$ can drop to $E_F$, the Fermi energy reaches the band edge $E_B$, where the curve must be truncated.  

To maximize the density range for superfluidity, the magnetic field $B$ should be made large.  In the ferromagnetic stripes shown in 
Fig.\ \ref{Fig.1}, the maximum magnetic field  is $B\simeq 1.2$ T, with magnetic length $\ell_B\sim 24$ nm.  For $\kappa=3$,  this corresponds to a minimum stripe width  
$d_B=d_{\textrm{min}}\ell_B=74$ nm, readily  attainable experimentally.  
However for $B\alt 1.2$ T, the magnetic band width  is narrow so that $E_F$ reaches $E_B$ 
at low densities (see Fig.\ \ref{Delta}). 

To obtain superfluidity at higher densities, 
a deformation of the graphene layer can be used to produce much larger pseudo-magnetic fields.   
Periodic deformations of the graphene layers generate a fictitious vector potential which can produce a periodic pseudo-magnetic field
in the layers \cite{Low2010,Guinea2008,Guinea2009a,guinea_gauge_2009}.  

The pseudo magnetic field induced by the strain changes
sign for the two valleys. The Dirac cones at the Brillouin
zone points $K$ and $K^\prime$ will experience an alternating
magnetic field  in both cases, but with a $\pi$ shift in  phase.  However, in the absence of 
intervalley scattering, as considered in this work, a global $\pi$ shift
in the zero-flux magnetic field does not affect the renormalization of
the Fermi velocity of the two Dirac cones.  Therefore, the evaluation procedure for determining the RPA screening and the 
density of carriers, using two equivalent renormalized Dirac cones, is the same as for the 
ferromagnetic stripes. 

\begin{figure}[t]
\includegraphics[height=0.45\textwidth,angle=-90]{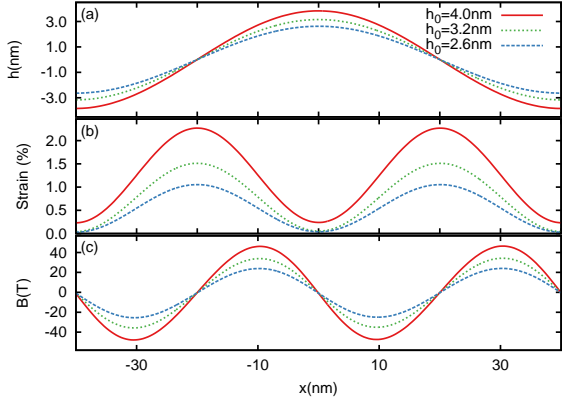}
\caption{(Color online) (a) Periodic deformation profile $h$ of a graphene sheet. (b) Induced strain in the sheet. (c) Induced  pseudo-magnetic field in the sheet.}
\label{fig1.LC}
\end{figure}

Let us consider a periodic modulation of the graphene layer with height 
profile [Fig.\ \ref{fig1.LC}(a)] along the zig-zag direction of the lattice, 
$h(x,y)=h_0\cos\left(2\pi x/ \lambda \right)$.
$h_0$ and $\lambda$ are the modulation amplitude and wavelength.  The  
pseudo-potential $\vec{A}=(A_x,A_y)$ at  lattice point $\alpha$ is defined as:
\begin{equation}
A_x^\alpha + i A_y^\alpha = \frac{1}{ev_F}\sum_{\beta \in nn(\alpha)} t_{\alpha\beta}(R_{\alpha\beta}) 
\mathrm{e}^{\left(-i \vec{K}\cdot\vec{R}_{\alpha\beta}\right)} \ .
\end{equation}
The $K$ point of the Brillouin zone is at $\vec{K}$, 
 $\vec{R}_{\alpha\beta}$ is the distance between atoms $\alpha$ and $\beta$, and the  
hopping amplitude $t_{\alpha\beta}$ couples the $p_z$ orbitals 
on  neighboring atoms. 
The corresponding pseudo-magnetic field is $\vec{B}=\vec{\nabla} \times \vec{A}$.

Figure \ref{fig1.LC}(a) shows the profile of the periodic deformation along the zig-zag direction for $\lambda=80$ nm {and amplitudes} $h_0$. 
Figure \ref{fig1.LC}(b) shows {the induced} strain, and Fig.\ \ref{fig1.LC}(c)  the induced periodic pseudo-magnetic field. 
The periodicity of the strain profile and the pseudo-magnetic fields is one-half of $\lambda$, so $\lambda=80$ nm corresponds to a 
stripe width $d_B=20$ nm.

To explore the modification of the electronic properties induced by an out-of-plane deformation, we have calculated the local density of states 
(LDOS) using the tight-binding Hamiltonian with a spatially varying hopping amplitude induced by a spatially varying inter-carbon distance.
The LDOS is calculated through a Chebyshev expansion of the single particle Green's function  \cite{Neek-Amal2012,Neek-Amal2013}.  
For $\lambda=80$ nm, we find that for deformation amplitudes up to $h_0 < 4$ nm,  we recover the same phenomenon in the energy dispersion as that 
for the real magnetic field, an  isotropic linear dispersion  with the slower Fermi velocity $\alpha_dv_F$.
Figure \ref{fig1.LC}(c) shows that an amplitude $h_0 \simeq 2.6$ nm  
generates the large pseudo-magnetic field $B\simeq 20$ T.  
This leads to a 
much wider magnetic band width $E_B$ than is possible for ferromagnetic stripes, and  $E_F$ does not reach $E_B$ until densities 
$n  > 10^{11}$ cm$^{-2}$. 
An experimental  realization would be to deposit graphene on a substrate that can be strained through surface acoustic 
waves (SAWs) \cite{Fu2010}, using two inter-digital transducers and a piezo-electric substrate.  
For wavelengths $\lambda\simeq 80$ nm and  typical piezo-electric materials used in SAW devices, the frequency 
$\sim 50$ GHz. 

The Fourier transform of the interaction  in Eq.\ (\ref{Vscr})  in the presence of graphene corrugations should be evaluated in 
the periodic curved geometry.  However, in the strong coupling regime,  
the electron-hole pairs have dimensions along the layers comparable to the deformation wave-lengths, so the curved geometry corrections to the interaction will be small.

\subsection{Transition temperature}

Over the range of parameters we are considering, we find that the electron-hole superfluidity  
is always in the strong-coupling regime, well inside the crossover regime of the 
BCS-BEC crossover.  The transition temperature calculated within the mean 
field approach, $T_{mf}$, will be much larger than the actual transition temperature for the onset of 
phase coherence.  A lower bound on the  transition 
temperature in  two-dimensions is given by 
the  Kosterlitz-Thouless (KT) temperature,
\begin{equation}
T_{KT} = (\pi/2)\rho_s(T_{KT})\ ,
\label{KostTh}
\end{equation}
where $\rho_s(T)$ is the superfluid stiffness.  
 $\rho_s(T=0)=E_F/4\pi$ and, when  
$k_F D$ is small, $\rho_s(T)$ falls off slowly with $T$ for $T\ll \Delta_{\mathrm{max}}$.  Hence taking $\rho_s(T)\simeq\rho_s(0)$, we  obtain  $T_{KT}= E_F/8$ for $E_F/8 \ll\Delta_{\mathrm{max}}$. 

Since in general $T_{KT}\ll T_{mf}$, the mean field superfluid gap will be insensitive to $T$  for 
$T\leq T_{KT}$.   By a similar argument, the RPA screening polarization bubbles are only weakly affected by finite $T\leq T_{KT}$.  
Thus, in the superfluid calculations we can take $T=0$.

\begin{figure}[h]
\includegraphics[height=4.9cm,width=0.48\textwidth]{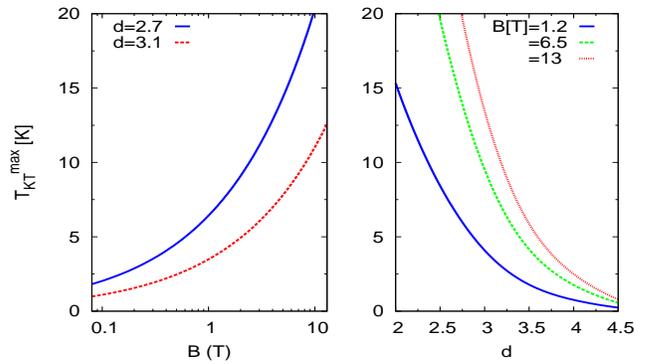}
\caption{(Color online) 
(a) $T_{KT}^{\textrm{max}}$ as a function of  $B$. 
Array spacings $d=3.1$ and $2.7$ give maximum transition temperature 
for embedded and free-standing system,  respectively.
(b) $T_{KT}^{\textrm{max}}$ as a function of $d$.}
\label{EFmax}
\end{figure}

For superfluidity to occur, we recall that the array spacing $d\agt 3$ (Fig.\ \ref{rs-alpha}). 
For these values of $d$, the linearized Eq.\ (\ref{epsilonk}) is valid whenever $E_F<E_B$ is satisfied.   This inequality 
establishes an upper bound on the maximum transition temperature, $T_{KT}\leq T_{KT}^{\textrm{max}}\simeq 35 \sqrt{B \mathrm{(Tesla)}} \exp{(-d^2/4)}$ K, the equality occurring for $E_F=E_B$. 
Figure \ref{EFmax} shows this maximum $T_{KT}^{\textrm{max}}$ as a function of both $B$ and $d$.  

\section{Conclusions}

We have proposed an electron-hole double monolayer graphene system, 
in which a quantum phase transition to a superfluid is induced by 
a periodic magnetic field. 
Electron-hole pairing in conventional double monolayers 
is known to be severely weakened 
by screening within the layers which kills the superfluidity.  
If the pairing can be made strong,  
then a large superfluid gap is known to open up, destroying the low-energy 
single-particle excitations that cause the screening. 
We show how a periodic magnetic field applied perpendicular to the monolayers 
could be used for this purpose: The field preserves the isotropic Dirac cones 
of the original monolayers but reduces the Fermi velocity in a tunable way, 
shifting the system parameters into the strongly coupled pairing regime 
where screening is sufficiently weakened for finite-temperature 
electron-hole superfluidity to occur. 

\begin{acknowledgments}
We thank M. Zarenia for useful discussions. 
L.D. acknowledges financial support from MIUR: FIRB 2012, 
Grant No.\ RBFR12NLNA\_002, and PRIN, Grant No.\ 2010LLKJBX. 
A.P. and D.N. acknowledge financial support from  University of Camerino  
FAR 2012 project CESEMN.  
L.C. acknowledges financial support from Flemish Science Foundation (FWO).

\end{acknowledgments}

\end{document}